\documentclass[11pt]{article}
\usepackage{graphicx}
\begin{document}
\nopagebreak[4]
\title{TCP Congestion Control Identification}
\author{Mohammed Ahmed}
\maketitle
\begin{abstract}
Transmission Control Protocol (TCP) carries most of the traffic on the Internet these days. There are several implementations of TCP, and the most important difference among them is their mechanism for controlling congestion. One of the methods for determining type of a TCP is active probing. Active probing considers a TCP implementation as a black box, sends different streams of data to the appropriate host. According to the response received from the host, it figures out the type of TCP version implemented.
   TCP Behavior Inference Tool (TBIT) is an implemented tool that uses active probing to check the running TCP on web servers. It can check several aspects of the running TCP including initial value of congestion window, congestion control algorithm, conformant congestion control, response to selective acknowledgment, response to Explicit Congestion Notification (ECN) and time wait duration. In this paper we focus on congestion control algorithm aspect of it, explain the mechanism used by TBIT and present the results. 
\end{abstract}
\section{INTRODUCTION}
   Transmission Control Protocol (TCP) is widely used on the Internet, and most of the traffics are carried by this protocol. So performance of the Internet depends significantly on TCP's performance. One of the most important aspects of controlling traffic is dealing with congestions. Since there are several types of TCP, each of them has implemented their own congestion control mechanism that is the main difference among them.  

   For doing some kinds of Internet researches, sometimes we need to know what type of TCP is running on a special host. Since we neither have access to the source code of it nor the permission to log into that host and check, we should find a way to check the TCP type remotely. In this case active probing will be useful \cite{bib1, bib10}. Active probing considers the remote host as a black box. It sends different streams of data to the remote host and the remote host responds to the stream. According to the response, we can infer what type of congestion control is used and consecutively what type of TCP is running. 

   Actually it is not possible to check arbitrary types of remote hosts. Because some of them may not respond to our request for several reasons like lack of permission, blocking the streams by the firewall etc. So we have to restrict our hosts to only web servers. It is obvious that they will always respond to our request for getting a page.  
   In this paper we use TCP Behavior Inference Tool (TBIT) that tests TCP Tahoe \cite{bib9}, Reno \cite{bib5} and NewReno \cite{bib7}. So we can only check the congestion control mechanisms on web servers supporting one of these TCP types \cite{bib2}.

  The remainder of this paper is organized as follows. Section two explains different congestion control mechanisms that are used nowadays. Section three describes a mechanism that will be used in order to distinguish the congestion control type that is running on a web server. Section four uses that mechanism in practice and shows the results of the test. Section five concludes everything in this paper and raises some future works.

\section{DIFFERENT CONGESTION CONTROL MECHANISMS}
There are different mechanisms for controlling congestion implemented. Congestion control consists of four parts: slow start, congestion avoidance, fast retransmit and fast recovery. The slow start and congestion avoidance algorithms must be used by TCP senders to control the amount of outstanding data being injected into the network.    For this reason, the sender starts with a low rate of sending data segments, and make the data amount twice after receiving acknowledgment (ACK) for each of them. After crossing a threshold it will increase the rate of sending data one by one.

   A TCP receiver should send an immediate duplicate ACK when an out-of-order segment arrives. The purpose of this ACK is to inform the sender that a segment was received out-of-order and which sequence number is expected. Then the sender enters fast retransmission and fast recovery stages and according to the TCP type that is used, the sender will change the transmission rate in order to control the traffic. The original definition of TCP congestion control mechanisms could be found in RFC-2851. In the following parts of this section we will briefly describe three different mechanisms implemented to control the congestion within Fast Retransmit and Fast Recovery stages.

2.1 Tahoe\\
  TCP Tahoe is one of the earliest implemented TCPs. It only contains Slow Start, Congestion Avoidance and Fast Retransmit. After receiving a small number of duplicate ACKs for the same TCP segment, the data sender infers that a packet has been lost and retransmits the packet without waiting for a transmission timer to expire, leading to higher channel utilization and connection throughput  \cite{bib3}. Then it goes to Slow-Start stage, and go on working from that point.

2.2 Reno\\
   The Reno TCP implementation retained the enhancements incorporated into Tahoe, but modified the Fast Retransmit operation to include Fast Recovery. The new algorithm prevents the communication path (pipe) from going empty after Fast Retransmit, thereby avoiding the need to Slow-Start to refill it after a single packet loss. Fast Recovery operates by assuming each duplicate ACK received represents a single packet having left the pipe. Thus, during Fast Recovery the TCP sender is able to make intelligent estimates of the amount of outstanding data.
   Fast Recovery is entered by a TCP sender after receiving an initial threshold of duplicate ACKs. This threshold, usually known as tcprexmtthresh, is generally set to three. Once the threshold of duplicate ACKs is received, the sender retransmits one packet and reduces its congestion window by one half. Instead of slow-starting, as is performed by a Tahoe TCP sender, the Reno sender uses additional incoming duplicate ACKs to clock subsequent outgoing packets  \cite{bib3}.

2.3 NewReno\\
   NewReno TCP includes a simple change to TCP Reno and improves its performance. The change concerns the sender' s behavior during Fast Recovery when a partial ACK is received that acknowledges some but not all of the packets that were outstanding at the start of that Fast Recovery period. In Reno, partial ACKs take TCP out of Fast Recovery by “deflating” the usable window back to the size of the congestion window. In NewReno, partial ACKs do not take TCP out of Fast Recovery. Instead, partial ACKs received during Fast Recovery are treated as an indication that the packet immediately following the acknowledged packet in the sequence space has been lost, and should be retransmitted. Thus, when multiple packets are lost from a single window of data, NewReno can recover without a retransmission timeout, retransmitting one lost packet per round-trip time until all of the lost packets from that window have been retransmitted. New-Reno remains in Fast Recovery until all of the data outstanding when Fast Recovery was initiated has been acknowledged  \cite{bib3}.

\section{AN ARCHITECTURE FOR IDENTIFYING CONGESTION CONTROL}
  Explaining briefly the architecture, a connection is established between two hosts and they start sending and receiving data segments like a regular TCP connection. Then some packets are dropped intentionally to check the behavior of remote host. Within the following parts of this section we explain these steps more comprehensively. 
   First of all we need a connection between our machine and the remote host. This connection will not be a regular connection, because we don't want the kernel to manipulate our streams. In the other words, we need an interface to connect directly to the remote host. We create our own TCP packets, and acknowledge only the desired ones. 
   BSD Packet Filter (BPF) is kernel agent that provides a user-level packet capture. It captures specific packets and discards others as soon as possible. BPF contains two components: the network tap and packet filter. Network tap collects copies of packets from the network device drivers and delivers them to the application. The filter decides if a packet should be accepted and, if so, how much of it to copy to the listening port  \cite{bib4}. 
   BPF is good choice to be used as an interface to the remote host. So we use raw sockets and fabricates TCP packets. Then the packets are sent to the network card to be sent over the network. In the opposite way, when a packet is received from the remote host, firewall will block it so it won't be received by the kernel. BPF makes a copy of it and delivers this packet to our software that is working in the user-level space. Then we can easily decide when to create an ACK and when to drop a packet. Let's take a look at an algorithm that can identify the congestion control mechanism, considering our machine is A and the remote host is B.
\begin{itemize}
\item Open a raw IP socket 
\item Open a BPF device and set the filter to capture all packets going to and coming from host B.
\item Set up a host firewall on A to prevent any packets coming from host B from reaching the kernel of host A.
\item Create and send a TCP SYN packet with the destination address of host B and a destination port of 80. The Maximum Segment Size (MSS) is set to be a small number. This will make the host send several packets for a regular request. 
\item 	Request the base web page
\item 	The remote server starts sending the base page with small packets.
\item 	Create ACKs according to TCP protocol
\item 	Drop some packets
\item 	Check what the response will be from the remote host
\item 	Close the connection after a short time
\end{itemize}
   Using this algorithm, our software will receive different responses from different TCPs implemented. Now the software should analyze to understand the type of TCP that creates such responses.

\section{PRACTICAL EXAMPLE AND RESULTS}
Now according to the algorithm described in previous section we would like to create a practical example and see the results of that. So we will follow the same algorithm with some specific features as follows.
\begin{itemize}
\item	Establish a connection between A and B according to what was explained in the algorithm.
\item	Send a TCP SYN packet with the destination address of host B and a destination port of 80. The Maximum Segment Size (MSS) is set to 100 bytes.
\item	Request the base web page
\item	The remote server starts sending the base page with  100-byte packets.
\item	Create ACKs according to TCP protocol up to packet 13.
\item	Drop packet 13.
\item	Receive and acknowledge packets 14 and 15, but the acknowledge will be three times for packet 13 that will make sender go to Fast Retransmission state.
\item	Drop the packet 16.
\item	Receive and acknowledge other packets up to 25
\item	Close the connection after acknowledging packet 25.
\end{itemize}

   Now according to responses received from the remote host, we can infer what type of TCP congestion control is running on that machine. As we explained in 2.3 NewReno is characterized by Fast Retransmission for packet 13, no additional Fast Retransmits or Transmit Timeouts, and no unnecessary retransmission of packet 17, as in figure 1. Like what was stated in 2.2 Reno TCP is characterized by Fast Retransmit for packet 13, a Transmit Timeout for packet 16 and no unnecessary retransmission for packet 17, as in figure 2. As in 2.1 Tahoe TCP is characterized by no Retransmit Timeout before the retransmission of packet 13, but an unnecessary retransmission of packet 17, as shown in figure 3  \cite{bib2}. Since TCP is a user-configurable transport protocol  \cite{bib8}, responses received from different hosts are not limited only to these three types. So we would face some other types of congestion controls that are not widely used.

   As an explanation square signs in these figures stand for packets and plus signs are acknowledges for specific packets.
   According to  \cite{bib2}, it has been tested on 3728 web servers and the summary of them is shown in Table 1. Even though the number of tests has been more than 3728, only this number of servers replied TBIT request. There are several reasons for this problem as follows.
\begin{itemize}
\item	TBIT may detect packet reordering.
\item	An Internal buffer overflowed. 
\item	Based on the packets received, the software couldn't classify it in one of the TCP classes.
\end{itemize}
   
   As an explanation to Table 1, we should mention that TCP without retransmission was expected to be obsolete, but we can see lots of them running on some web servers. And also a number of web servers have implemented a variant of Reno TCP that transmits additional packets between retransmission of packets 13 and 16 that  \cite{bib2} called them Reno-Plus.

\section{CONCLUSION AND FUTURE WORKS}
In this paper we explained one mechanism for testing the congestion control running on a web server. This mechanism has been implemented and tested in TBIT \cite{bib2}. The source code and documentations are available on its web site  \cite{bib6}. This tool has been developed on FreeBSD platform, and can partially work on Unix. 
 Using this software we can test several TCP implementations without having access to their source codes. This might also be used for debugging TCP implementations. One other usage of this software may be making some statistics about the traffic on Internet that will be useful in developing softwares trying to shape or monitoring the traffic. 
  As a future work, someone can implement it on other transport protocols supporting congestion control like DCCP and SCTP. Since they are using congestion control algorithms somewhat like TCP, this mechanism can be helpful for them as much as it is for TCP. All the mechanisms will be the same except the part for creating TCP packets. In each of these two cases the software should create its own packet type i.e. DCCP or SCTP packet. Since they don't carry the traffic more than TCP, it may not be used very often on the Internet.

\end{document}